# Security Testing Framework for Web Applications: Benchmarking ZAP V2.12.0 and V2.13.0 by OWASP as an example

Usha-Sri Potti[1], Hsuan-Tung Chen[2], and Hung-Min Sun[3,*]

[1]Institute of Information Security, National Tsing Hua University, Hsinchu, Taiwan

[2]Industrial Technology Research Institute, Hsinchu, Taiwan

[3,*]Department of Computer Science, National Tsing Hua University, Hsinchu, Taiwan

[*]Email: hmsun@cs.nthu.edu.tw

## Abstract

The Huge growth in the usage of web applications has raised concerns regarding their security vulnerabilities, which in turn pushes toward robust security testing tools. This study compares OWASP ZAP, the leading open-source web application vulnerability scanner, across its two most recent iterations. While comparing their performance to the OWASP Benchmark, the study evaluates their efficiency in spotting vulnerabilities in the purposefully vulnerable application, OWASP Benchmark project.

The research methodology involves conducting systematic scans of OWASP Benchmark using both v2.12.0 and v2.13.0 of OWASP ZAP. The OWASP Benchmark provides a standardized framework to evaluate the scanner's abilities in identifying security flaws, Insecure Cookies, Path traversal, SQL injection, and more. Results obtained from this benchmark comparison offer valuable insights into the strengths and weaknesses of each version of the tool. This study aids in web application security testing by shedding light on how well-known scanners work at spotting vulnerabilities. The knowledge gained from this study can assist security professionals and developers in making informed decisions to support their web application security status. In conclusion, this study comprehensively analyzes ZAP's capabilities in detecting security flaws using OWASP Benchmark v1.2. The findings add to the continuing debates about online application security tools and establish the framework for future studies and developments in the research field of web application security testing.

Keywords: Computer Science; Security Testing; OWASP

## Introduction

Modern technology is significantly influenced by the internet, a web of links that has revolutionized the digital age. It has reshaped how we interact, do business, and access information. The development of artificial intelligence (AI) at an exponential rate is one of the main forces causing this transition. Due to this development in AI, the usage of Web Applications has been drastically increased. These web applications are commonplace in a number of industries, including finance, the military, banking, and retail.

With great innovation comes significant responsibility, particularly in the digital realm. The proliferation of web applications has raised profound security concerns, particularly while handling sensitive data. It's not just about delivering content anymore; it's about safeguarding the integrity and security of data. These security concerns encompass Confidentiality, Integrity, and Availability (CIA). These principles constitute the CIA triad, protecting data and applications against unauthorized access, malicious attacks, and service disruptions.

TABLE I. CIA Triad

| | |
|---|---|
| Confidentiality | Guarantees that private, sensitive data is only accessible to those who are permitted. |
| Integrity | Guarantees that data remains unaltered and trustworthy throughout its lifecycle. |
| Availability | Guarantees that web applications are available and working properly as of requirement without interruption. |

Ensuring all the CIA triad's principles is a tedious task. The complexity of modern web applications and the ever-evolving landscape of cyber threats increase the complexity to safeguard these critical aspects of data and application security.

Here, benchmarks like OWASP (Open Web Application Security Project) come into play. OWASP provides a comprehensive framework for identifying, understanding, and mitigating web application vulnerabilities. It offers a structured approach to addressing security concerns within web applications, helping developers and security professionals stay one step ahead of potential threats.

One essential tool in the arsenal of securing web applications is web application vulnerability scanners. These sophisticated software solutions systematically crawl through web applications, identifying vulnerabilities that may compromise the CIA triad. They can detect vulnerabilities, from common flaws like Cross-Site Scripting (XSS) and SQL Injection to more intricate security issues.[1]

### A. Background

Penetration Testing or pen testing, systematically assesses computer systems, networks, or web applications to recognize vulnerabilities that malicious actors would potentially exploit. It focuses on simulating real-world attacks in a controlled environment with the aim of evaluating the security status of the target system. This method is widely employed and highly effective for identifying potential threats or vulnerabilities in a web application.[2]

Most prominent approaches for pen testing are:

- **DAST:** In this approach, the tester, often without prior knowledge of the target system, simulates an attack from an external perspective, much like a malicious hacker would. Black box testing aids in identifying vulnerabilities that are visible or exploitable from outside.
- **SAST:** Whereas, in whitebox testing, the tester has prior knowledge of the target system, encompassing application's source code, design and architecture. This approach allows an in-depth assessment of the system, focusing on vulnerabilities that might not be apparent from an external perspective.

Based on Black Box and White Box Testing techniques, organizations use the specific tools and methods which include Dynamic Application Security Testing (DAST) and Static Application Security Testing (SAST).

- **DAST:** A black box testing technique that evaluates web applications through an external perspective. It simulates real-world attacks by interacting with the running application to identify vulnerabilities, such as configuration issues or security misconfigurations. Organizations typically employ DAST during the testing phase to assess the application as a whole, similar to how a potential attacker would externally interact with it.

- **SAST:** A white box testing technique that evaluates the source code, binaries, bytecode of the application to locate potential security problems by scanning for flaws and unsafe coding techniques. It is often used during the development phase to identify vulnerabilities early in the software development lifecycle.

For this study, the DAST tool OWASP ZAP is employed. Subsequently, the advantages and disadvantages of both the SAST and DAST techniques, along with their corresponding tools, will be explained and justified.

## B. Motivation

The critical importance of web application security in today's digital landscape drives this study. Web applications have become integral to various sectors, including finance, e-commerce, healthcare, and more. However, this widespread adoption also brings forth security challenges. Ensuring confidentiality, integrity, and availability within these applications is a pressing concern.

In the realm of web application security, knowledge is power. Understanding potential vulnerabilities and threats is crucial to safeguarding sensitive data and maintaining trust. This study aims to provide a deep understanding of web application security and assess the effectiveness of OWASP Zed Attack Proxy (ZAP), a renowned open-source Dynamic Application Security Testing (DAST) tool.

The focus on OWASP ZAP is significant. Among numerous DAST tools available, ZAP stands out as a robust choice for organizations seeking to enhance web application security. By evaluating the performance of OWASP ZAP versions 2.12 and 2.13 through a systematic comparative analysis, this study aims to provide valuable insights.

This study contributes to the field of web application security but also highlights the importance of benchmarking as an evaluation technique. Ultimately, this research empowers organizations with the knowledge to strengthen security postures in an ever-evolving digital landscape.

## C. Justification

Reason behind choosing DAST rather than SAST for this study:

- **External Perspective Assessment:** DAST concentrates on assessing web applications from the outside, simulating how actual attackers would interact with the program. This strategy is advantageous since it mimics how possible attackers would approach the system, finding weaknesses that would be exploited externally. SAST, on the other hand, evaluates flaws in the source code itself and might not fully capture scenarios of external threat.
- **Realistic Threat Simulation:** DAST tools, such as OWASP ZAP, simulate actual attacks, making them very effective in spotting flaws that potentially dangerous actors could exploit. This realism in threat simulation provides a more accurate understanding of the application's security status under real attack scenarios.
- **Testing the Running Application:** DAST can interact with the Web Application in runtime time in order to evaluate its behavior and responses in real-time. This dynamic assessment ensures that the evaluation reflects the application's current state, including configurations and runtime conditions. However, SAST only examines the static codebase, therefore it might miss runtime problems and configuration vulnerabilities.
- **Efficiency in Identifying Configuration Errors:** DAST is suitable to identify the errors in configuration or presence of security misconfigurations which leave the application vulnerable. These issues might be challenging to recognize through SAST alone, as they often involve runtime settings and environmental factors.

- **Black Box Testing for Realism:** DAST closely aligns with black box testing principles, which emulate an external, attacker-centric perspective. This choice is instrumental to assess the security of web applications against potential external threats realistically. It ensures whether the evaluation considers the application's exposure to the external environment.
- **Holistic Assessment:** DAST evaluates the application as a whole, examining all its components and interactions. This holistic assessment is crucial for identifying vulnerabilities that may arise from the integration of various elements within the application, such as APIs, databases, and third-party components.
- **Comprehensive Security Status:** DAST aims comprehensive evaluation of the web application's security concerns against harmful external threats. Through the identification of vulnerabilities across numerous attack channels, this method offers a more thorough understanding of security.

To sum up, the necessity for an external, practical perspective on security assessment drove the use of DAST approaches, with OWASP ZAP receiving special attention. This decision fits in perfectly with the main goals of the study, which are to thoroughly and effectively assess the security of online applications against external threats.

DAST stands out for its ability to emulate real-world attacks, assess runtime conditions, pinpoint configuration errors, and offer a holistic view of security. This strategic selection ensures that potential vulnerabilities, especially those readily exploitable by external adversaries, undergo comprehensive evaluation and remediation.

Adapting DAST techniques enhances security assessments, effectively addressing vulnerabilities and strengthening web application security against external threats.

## Related Work

Table II displays the previous comparitive studies based on the web application vulnerability scanners.

TABLE II. Previous comparative studies

| Paper | Scanner | Remarks |
|---|---|---|
| Abdullah, Himli S. [3] | Paros 3.2.13, 2.7.0. | Compared using Damn Vulnerable Web Application (DVWA) and Buggy Web Application. |
| Ahamed, Azaz [4] | Netsparker, Burp suite, ZAP. | Compared using top 5 Government Web Applications of different sectors based on OWASP top 10 (2017) vulnerabilities. |
| Al Anhar, Azwar [5] | ZAP 2.10.0, Wapiti 3.0.4, Burp suite 2021.4.3. | Compared using Damn Vulnerable NodeJS Application (DVWA) (OWASP top 2017) and Node Goat (OWASP top 2013). |
| Albahar, Marwan [6] | ZAP 2.11.0, Burp suite 2021.9.1, Qualys WAS 8.12.55-1, Arachni 1.5.1, Wapiti 3.0.5, Fortify Webinspect 21.2.0. | comparison based on the OWASP Top 10 vulnerabilities by manually benchmarking with metrics like coverage of crawling and more. |
| Amankwah, | Acunetix, HP | Comparison using Damn Vulnerable WEB |
| Richard [7] | Webinspect, IBM App scan, ZAP, Skipfish, Arachni, Vega, Iron WASP. | Application (DVWA) based on OWASP Web Benchmark Evaluation (WBE) and Web Application Security Scanner Evaluation Criteria (WASSEC). |
| Dasun, K. L. [8] | SAST: FBw FindSecBugs 1.4.6, FindBugs 3.0.1, SonarQube 3.14, PMD 5.2.3, DAST: ZAP vD-2016-09-05. | compared using the OWASP Benchmark project based on the Vulnerabilities and Metrics involved. |
| Di Stasio, Vincenzo [9] | 16 SAST tools | compared using the OWASP Benchmark project based on the Vulnerabilities and Metrics involved. |
| El, Malaka [10] | Nessus, Burp suite | Compared based on Web Application Vulnerability Scanner Evaluation Project (WAVSEP) Benchmark, Web Input Vector Extractor Teaser (WIVET) Benchmark and Scalability. |
| Higuera, Juan R. Bermejo [11] | 7 SAST tools: Fortify, Xanitizer, Findsecbugs, Klokwork, checkmarx, Coverity, Spotbugs | compared using the OWASP Benchmark project based on the OWASP top 10 Vulnerabilities. |
| Kalaani, Christopher [12] | ZAP, Snort | Compared based on various types of SQL injection. |
| Lavens, Emma [13] | 24 Vulnerability Scanners | Compared using Open Worldwide Application Security Project (OWASP) and Web Application Vulnerability Scanner Evaluation Project (WAVSEP) Benchmarks based on 11 vulnerabilities. |
| Makino, Yuma [14] | ZAP, Skipfish | Compared using Damn Vulnerable Web Application (DVWA) based on Web Application Vulnerability Scanner Evaluation Project (WAVSEP) Benchmark. |
| Mburano, Balume [15] | ZAP, Arachni | Compared using Open Worldwide Application Security Project (OWASP) and Web Application Vulnerability Scanner Evaluation Project (WAVSEP) Benchmarks for Command injection, SQL injection, Cross Site Scripting (XSS) and Path traversal vulnerabilities. |

## Methodology

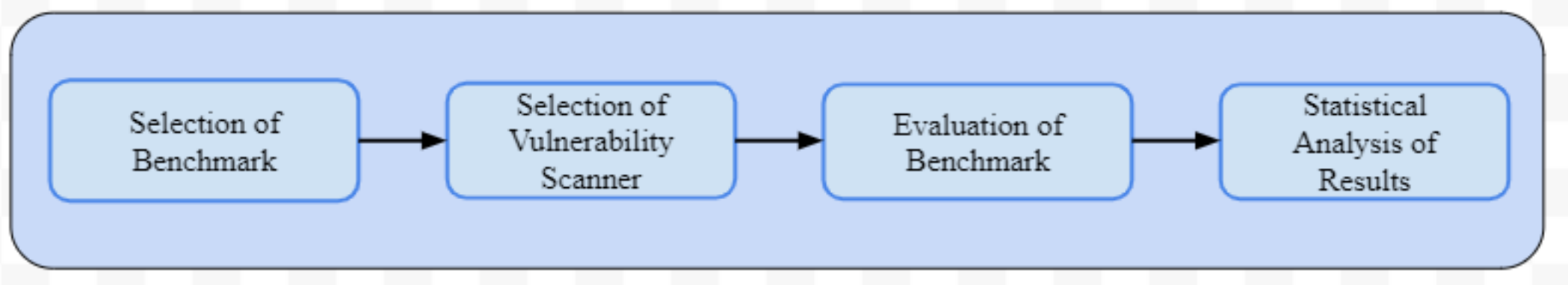

Figure 1. Methodology

A. Selection of Benchmark

The study relies on the OWASP Benchmark to evaluate two versions of OWASP ZAP, a DAST tool. This choice is strategic as the OWASP Benchmark specifically assesses web application security tools, aligning precisely with this study's focus. Additionally, it grounds itself in real-world attack scenarios, ensuring relevance to prevalent web application threats. The widespread recognition and acceptance of OWASP Benchmark within both the cybersecurity industry and the OWASP community bolster the credibility of this research. Furthermore, its open-source nature ensures transparency in methodologies and results, contributing to the trustworthiness of the study. The benchmark's continuous updates and alignment with industry standards underscore its relevance, making it an ideal choice for comparing different versions of OWASP ZAP. This decision aims to deliver a comprehensive evaluation of web application security, supported by OWASP Benchmark's comprehensiveness, relevance, recognition, and adherence to standards.

B. Selection of Vulnerability Scanner

Among various available DAST tools, OWASP ZAP emerges as the preferred choice due to its diverse capabilities and strong alignment with real-world threats. Widely adopted in the cybersecurity community, OWASP ZAP effectively identifies web application vulnerabilities and boasts a user-friendly interface suitable for both professionals and newcomers in security testing. Its open-source nature, coupled with a vibrant community, ensures continuous updates and adaptability to emerging threats. Aligning closely with OWASP standards, it guarantees reliability and comprehensive security assessments. The tool's extensive testing options, detailed reporting, and customization capabilities empower researchers to tailor evaluations according to specific web application needs. This strategic selection of OWASP ZAP aims to conduct a thorough evaluation of web application security against external threats, providing valuable insights to the cybersecurity community and enabling effective comparisons between different versions of the tool.

C. Evaluation of Benchmark

The OWASP Benchmark, a part of the diverse OWASP projects, stands significant, updated roughly every three years to align with prevalent vulnerabilities in the field. It involves various categories of attacks, each requiring unique mitigation strategies:

**Command Injection:** Attackers exploit improper command inputs, potentially leading to severe consequences. Strategies involve error handling, security updates, implementing a Web Application Firewall (WAF), and monitoring/logging.

**Path Traversal:** This attack aims to access restricted resources outside the webroot. Exploitation involves traversal attempts and accessing sensitive files, leading to data exposure or system compromise. Mitigation strategies include access controls, resource isolation, error handling, security updates, and implementing security headers.

**Secure Cookie Flag:** Failure to set the "Secure" flag on cookies makes them susceptible to interception, risking sensitive data exposure. Exploitation involves interception and unauthorized access to session cookies. Strategies include setting the "Secure" flag, using

HTTPS, employing HTTP Strict Transport Security (HSTS), and leveraging cookie attributes.

**SQL Injection:** This vulnerability arises from insufficient user input validation, allowing attackers to manipulate databases. Mitigation involves parameterized statements, input validation, Web Application Firewalls (WAFs), least privilege, error handling, security testing, database security, and developer education.

**XSS (Cross-Site Scripting):** Improper content validation before rendering in browsers leads to execution of malicious scripts. Exploitation and potential consequences involve data theft, defacement, or unauthorized tasks. Mitigation strategies include input validation, output encoding, Content Security Policy (CSP), sanitization libraries, regular security audits, developer education, and browser security updates.

Each vulnerability poses specific risks and requires tailored mitigation strategies to ensure web application security.

D. Statistical Analysis of Results:

To quantitatively evaluate ZAP's performance using the OWASP Benchmark, fundamental metrics for evaluation involve True Positives (TP), True Negatives (TN), False Positives (FP), and False Negatives (FN).

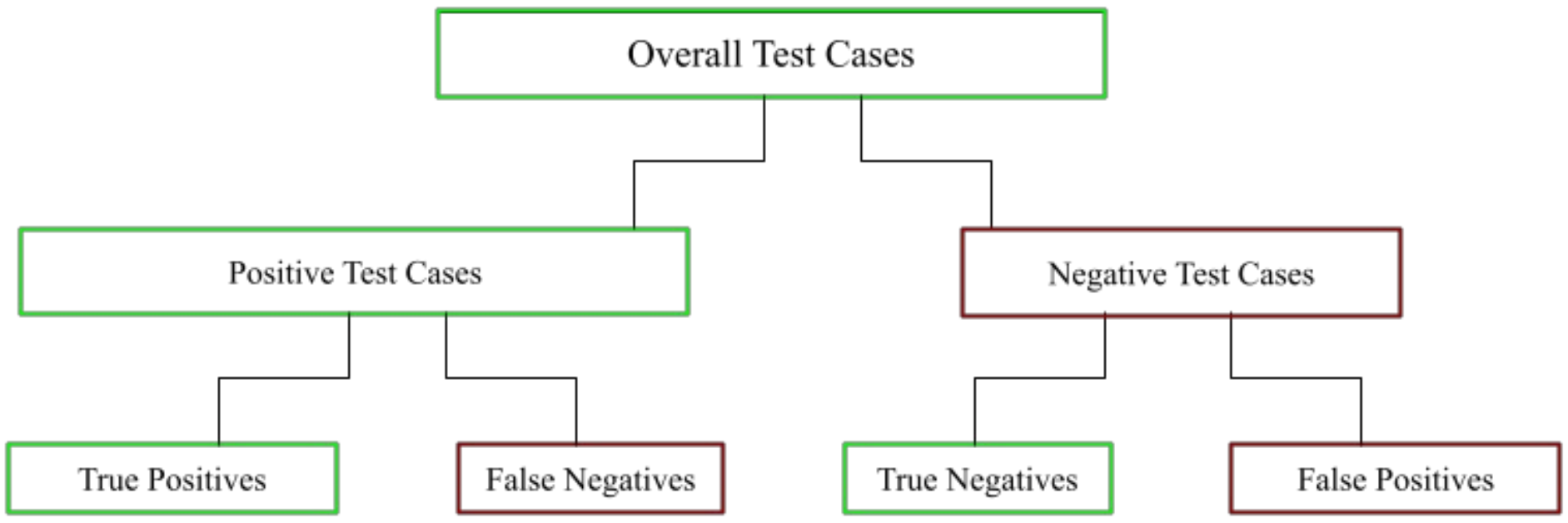

Figure 2. Summary of Evaluation Metrics

**True Positives (TP):** This metric involves the number of instances where the scanner accurately identifies the presence of an actual vulnerability in the web application within the OWASP Benchmark Project and labels them as real vulnerabilities.

**True Negatives (TN):** This metric involves the number of instances where the scanner accurately identifies the absence of vulnerability in the web application within the OWASP Benchmark Project and labels them as non-vulnerable.

**False Positives (FP):** This metric involves the number of instances where the scanner mistakes the presence of a vulnerability in the web application within the OWASP Benchmark Project and labels them as actual vulnerabilities.

**False Negatives (FN):** This metric involves the number of instances where the scanner fails to identify the presence of an actual vulnerability in the web application within the OWASP Benchmark Project and labels them as non-vulnerable.

These fundamental metrics, namely TP, TN, FP, and FN serve as foundational building blocks for metrics such as Accuracy, Recall, Precision, F1-score, and Youden Index. These supplementary metrics offer a deeper and more thorough insight of the scanner's functionality and its capacity to find vulnerabilities in the OWASP Benchmark.

**Accuracy:** These metrics find extensive usage in various fields, including Machine Learning and Data Sciences, Medical Diagnostics, Information Retrieval, and more. It is the ratio of correctly classified instances to the total number of cases in the dataset. In the context of the performance evaluation of ZAP, accuracy reflects the ability of ZAP to detect the presence of vulnerability among the actual vulnerabilities present in the OWASP Benchmark. [16]

$$Accuracy = \frac{correctly\ \ classified\ \ instances}{number\ \ of\ \ instances} \quad (1)$$

Were,

correctly classified instances=TP+TN (2)

Total number of instances=TP+TN+FP+FN (3)[17][18]

There is a limitation in using accuracy; it cannot distinguish between correct labels of different classes. Here the concepts of sensitivity and specificity become relevant.

**True Positive Rate (TPR):** This metric calculates the ratio of True Positive instances (TP) to the total number of actual positive cases in the dataset. When assessing ZAP's performance, TPR indicates its capacity to accurately detect vulnerabilities among the real vulnerability cases within the OWASP Benchmark.

were,

$$TPR = \frac{TP}{TP + FN} \quad (4)$$

Recall=Sensitivity=TPR (5)

**False Positive Rate (FPR):** This metric calculates the ratio of number of False Positive instances (FP) to the total number of actual negative cases in the dataset. When assessing ZAP's performance, this metric can determine specificity. Specificity demonstrates ZAP's ability to distinguish actual negatives (non-vulnerable cases) from the predicted negatives labeled as non-vulnerable.

were,

$$FPR = \frac{FP}{FP + TN} \quad (6)$$

Specificity= 1 - FPR (7)[19]

**Precision:** This metric calculates the ratio of number of True Positive instances (TP) to the total number of cases labeled true by the scanner. When assessing ZAP's performance, this metric indicates whether the identified positive cases are actual vulnerabilities among the predicted positive instances.

were,

$$Precision = \frac{TP}{TP + FP} \quad (8)[19]$$

**Youden's index:** This is a single-valued metric that combines both sensitivity and specificity to provide an overall measure of a model's performance. It aims to find the threshold that maximizes the difference between true positive rate and false positive rate.

were,

$$\text{Youden's index} = (\text{sensitivity} + \text{specificity}) - 1 \quad (9)[20]$$

E. Experimental Setup

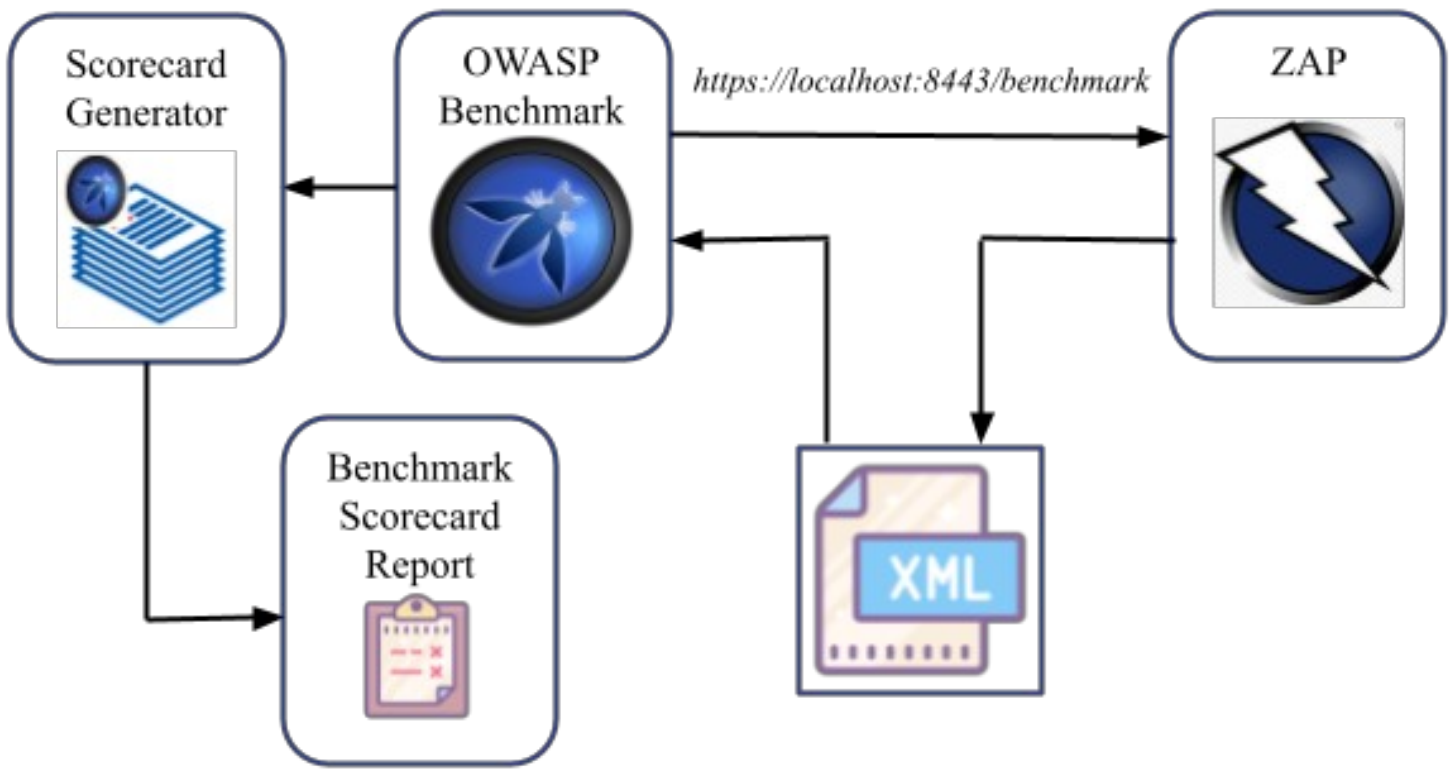

Figure 3. Experimental Setup

The experimental setup for this study primarily revolves around the OWASP Benchmark, a comprehensive assessment tool. This Benchmark includes a Scorecard Generator, developed using the Java programming language and a bash script. The key component of the Benchmark is its website, which encompasses 11 distinct vulnerability categories. When the Benchmark is executed locally on the host system, the interface, as depicted in the Figure 4, becomes accessible.

Each of these vulnerability categories comprises a set of test cases specifically designed to evaluate the presence of that particular vulnerability. These test cases are intended to be executed either against a website or within a vulnerability scanner, allowing for a thorough assessment of web application security.

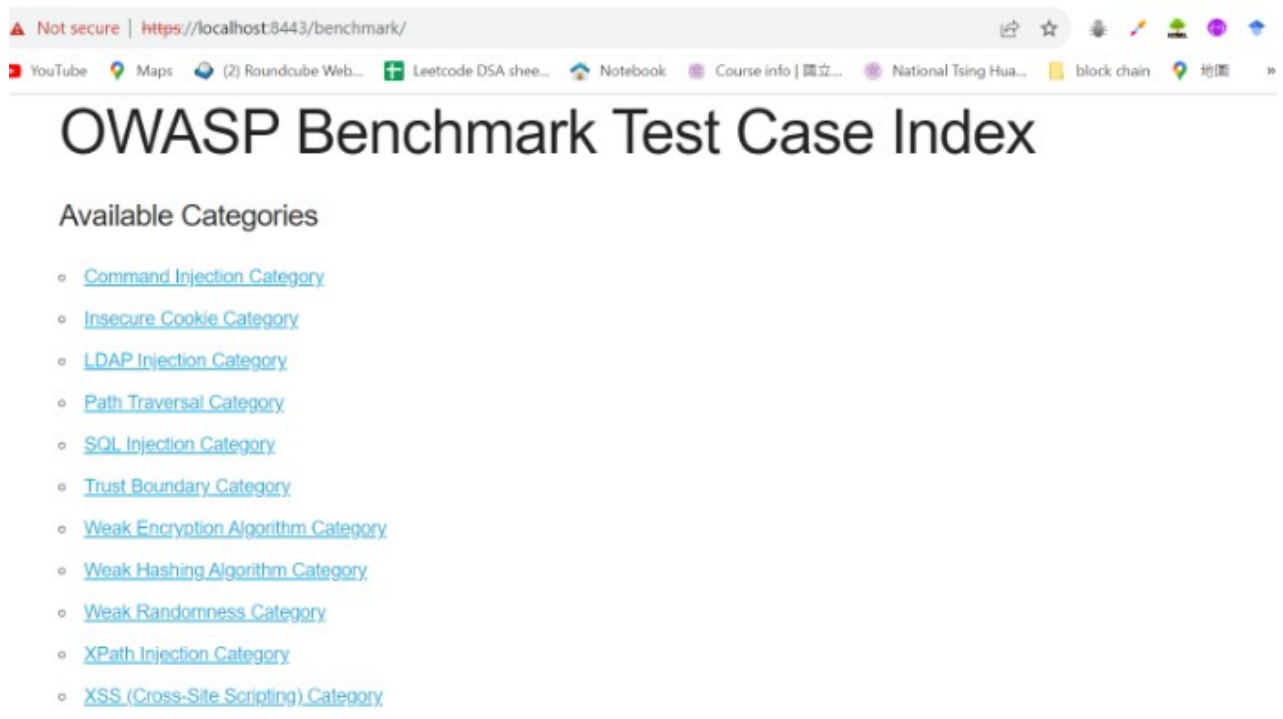

Figure 4. OWASP Benchmark-Vulnerability categories

Upon executing the Benchmark on the local host, the command prompt will display an interface similar to the one depicted in the Figure 5. A successful build will be indicated, and you can access the corresponding website on the local host via port 8443. This interface provides a user-friendly platform for initiating assessments and interacting with the OWASP Benchmark.

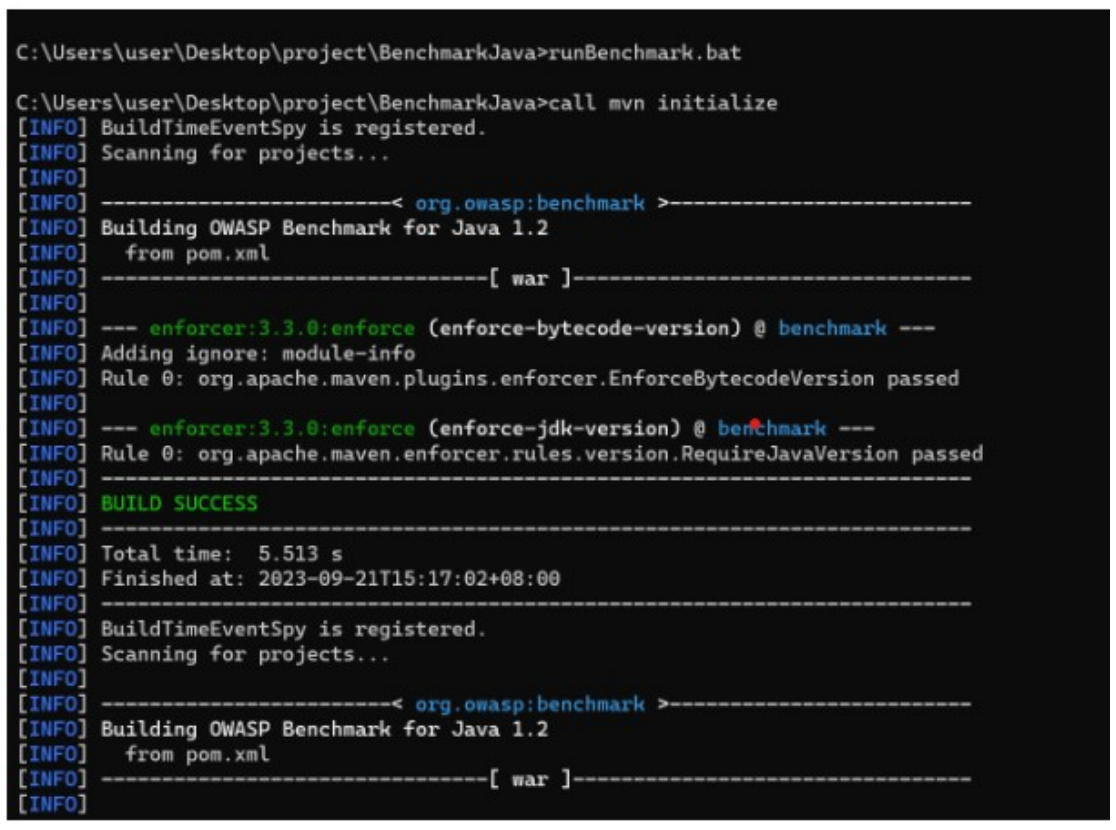

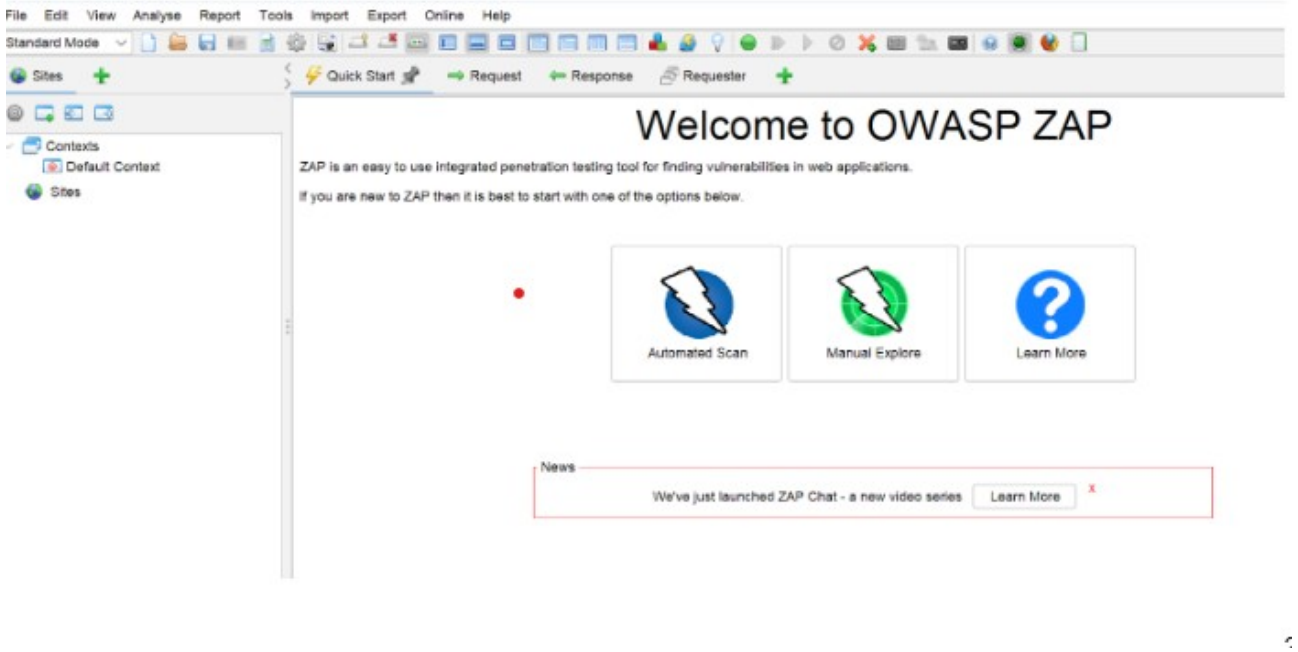

Figure 5. Console - Benchmark Execution Figure 6. OWASP ZAP Interface

Once the Benchmark has been executed on the local host, typically on port 8443, it is ready to undergo testing with vulnerability scanners, such as OWASP ZAP. These scanners are designed to identify vulnerabilities within web applications. The OWASP ZAP software interface, as depicted in the Figure 6, offers various functionalities, including an Automated Scan feature. This feature enables users to scan websites for vulnerabilities, provided they have the necessary permissions.

In this study, the Automated Scan feature is utilized to assess the OWASP Benchmark, which comprises sets of test cases for each vulnerability. The configuration for the Automated Scan, used in this research, is illustrated in the Figure 7.

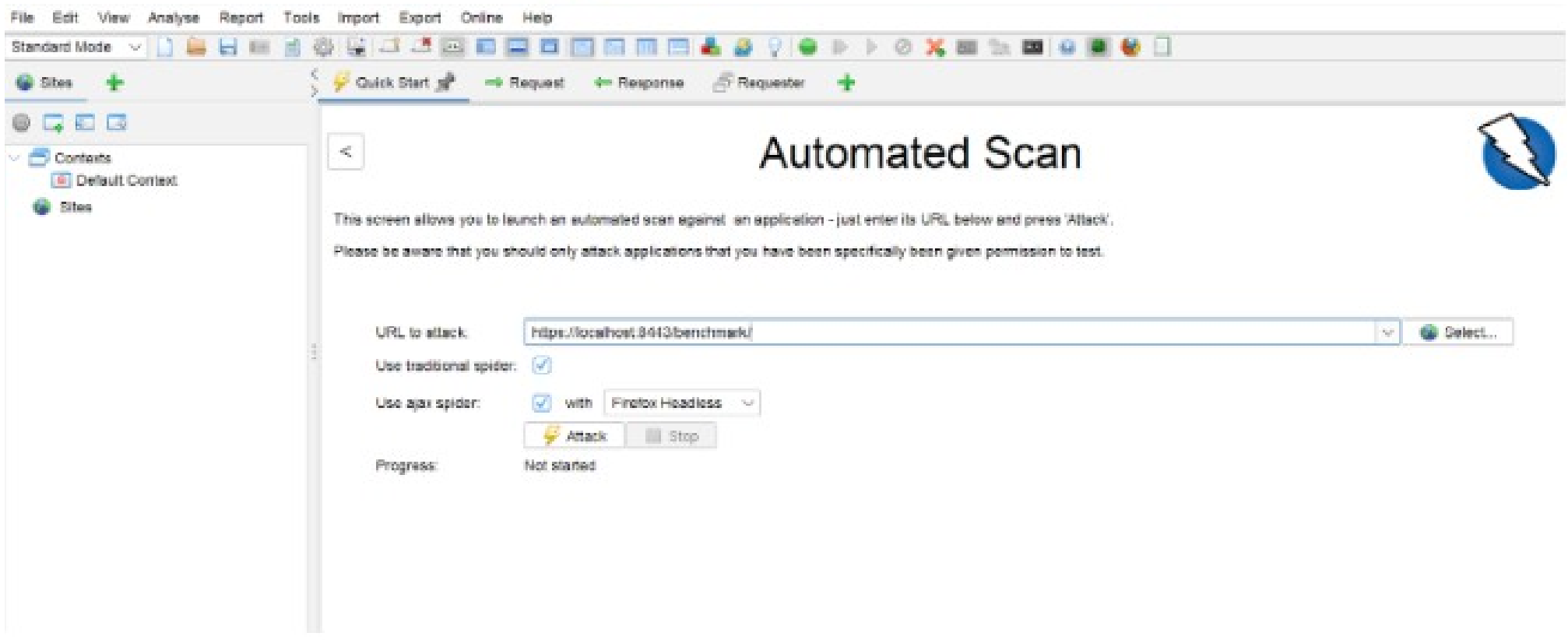

Figure 7. Automated Scan - Configuration

Once the scan begins, the scanner initiates a comprehensive crawl of the target website as shown in the Figure 8 using the web spider, systematically gathering all related web pages within the web application.

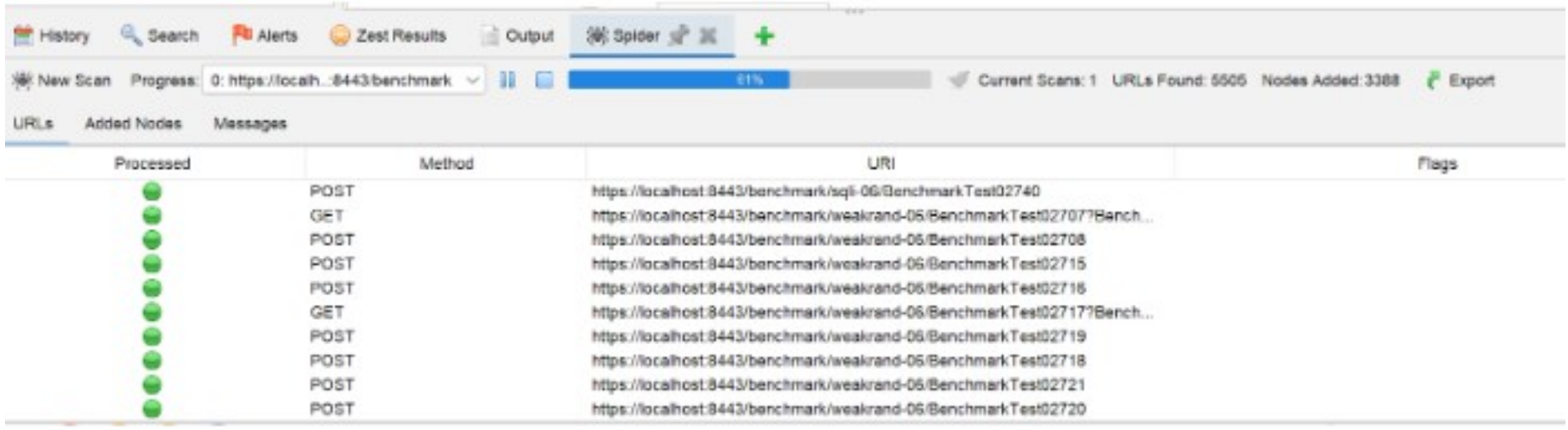

Figure 8. Web Spider

Subsequently, the active scanning phase commences as shown in the Figure 9, where the scanner searches for vulnerabilities and conducts assessments. These assessments play a crucial role in this analysis. Alerts generated during this scanning process are categorized in terms of risk, including High, Low, and Informational, providing a comprehensive view of potential vulnerabilities.

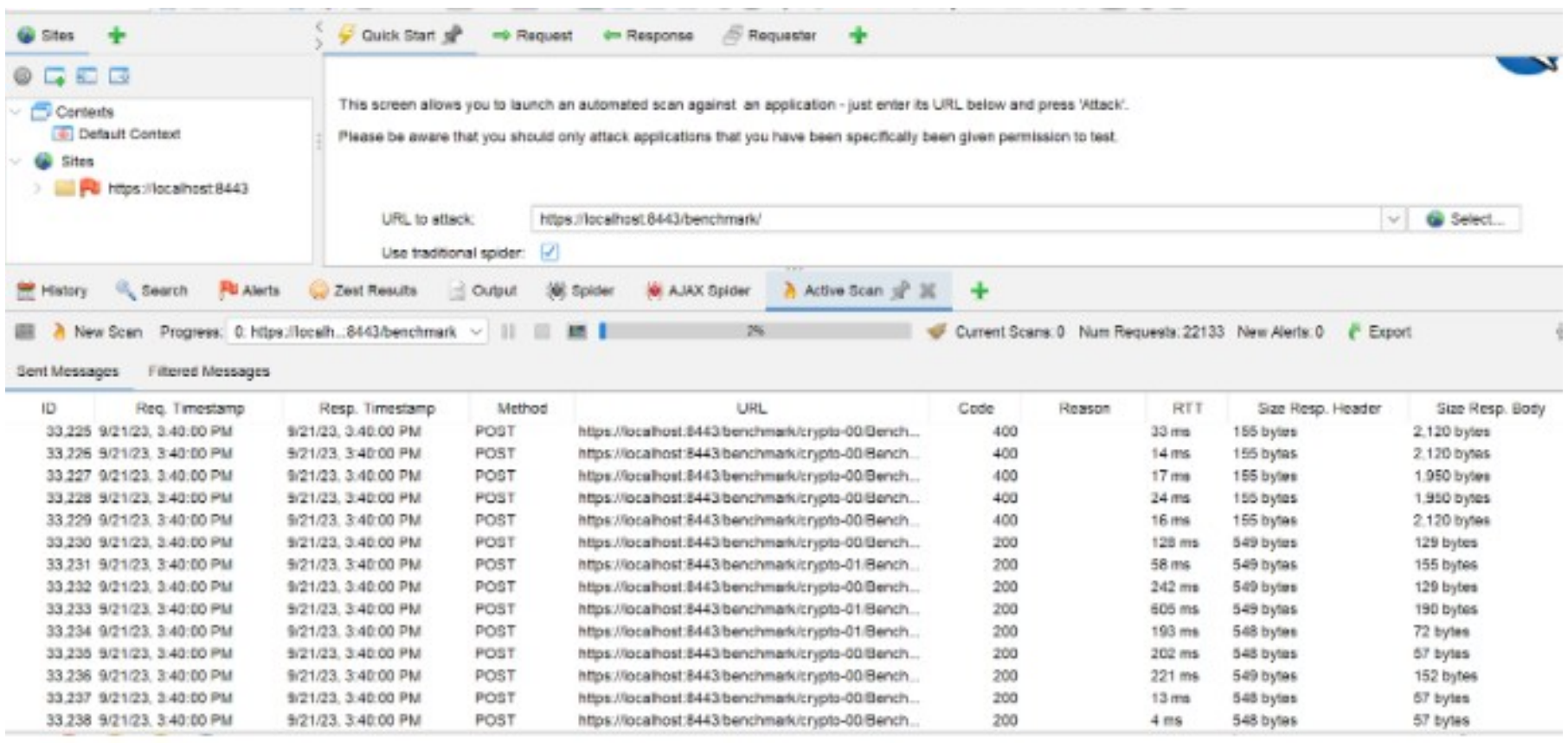

Figure 9. Automated scan process

After the completion of the scanning process, export the scan results in the form of a traditional XML Report. This report is helpful in generating the scorecard of the vulnerabilities present in the web application, that is OWASP Benchmark in this study. Scorecard generated by OWASP Benchmark gives valuable insights regarding the security assessment of the Vulnerability scanner, OWASP ZAP.

The interpretation of the Benchmark scorecard results is depicted in the Figure 10. This crucial step aids in comprehending and evaluating the security posture of the web application, highlighting potential vulnerabilities and their associated risk levels. The graph shows the comparison of vulnerability detection of the scanner against benchmark by analyzing the results using the metrics - Youden’s index.

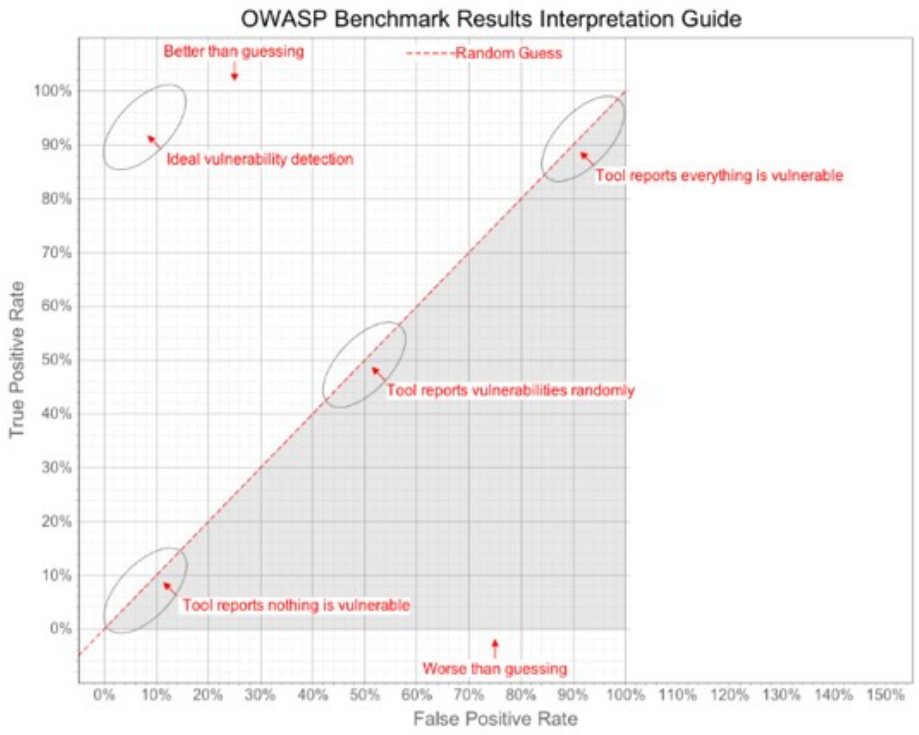

Figure 10. OWASP Benchmark Interpretation Guide [21]

**Comparative Analysis Results**

OWASP Benchmark’s examination of OWASP ZAP versions 2.12.0 and 2.13.0 has shed important light on the efficiency of application security testing tools across a number of vulnerability categories.

A. Analysis and comparison based on Precision.

TABLE III. Comparision of precision between v2.12.0 and v2.13.0

| Vulnerability Area | Precision | |
|---|---|---|
| | V2.12.0 | V2.13.0 |
| Command Injection | 100% | 83% |
| Path traversal | 100% | 100% |
| Secure Cookie Flag | 100% | 100% |
| SQL Injection | 99% | 96% |
| XSS (Cross-Site Scripting) | 100% | 100% |

Prior to diving into the thorough analysis provided by Youden's Index, it's critical to take into account precision metrics, which provide insightful information on how well OWASP ZAP versions are able to identify vulnerabilities.

- **Command Injection:** In version 2.12.0, precision stands at 100%, indicating that when this version generates an alert for command injection, it is almost always accurate. In version 2.13.0, the precision remains notably high at 83%, reinforcing the reliability of alerts.
- **Path Traversal:** Both versions exhibit perfect precision in identifying path traversal vulnerabilities, ensuring that the alerts raised are consistently accurate.
- **Secure Cookie Flag:** Similar to path traversal, both versions demonstrate perfect precision in detecting secure cookie flag vulnerabilities, further establishing the high accuracy of alerts.
- **SQL Injection:** While version 2.13.0 displays slightly lower precision compared to version 2.12.0 (96% vs. 99%), both versions maintain a high level of precision, affirming the reliability of generated alerts.
- **XSS (Cross-Site Scripting):** Both versions achieve a precision score of 100% in detecting XSS vulnerabilities, underscoring the trustworthiness of alerts.

A core understanding of alert reliability is provided by precision metrics. High precision ensures that when OWASP ZAP raises an alert, it is highly likely to be a true positive. These findings demonstrate the tool's effectiveness in reducing false positives, a crucial component of effective security assessments. When organizations use OWASP ZAP, they can depend on the alerts that are produced as accurate indicators of potential vulnerabilities.

While precision measurements show how reliable alerts are, it's important to supplement this analysis with Youden's Index, a more thorough evaluation tool that takes false negatives into account. In the subsequent section, we will delve into Youden's Index, providing a more holistic view of OWASP ZAP's performance in identifying vulnerabilities. This dual strategy helps enterprises choose the best version for their unique security requirements based on information that is available, which advances the field of web application security.

B. Analysis based on Youden's index.

**1) Command Injection:**

- **Comparison of Results:** Table IV is used to compare the capability of detecting the command injection vulnerability between the versions of ZAP v2.12.0 and ZAP v2.13.0 in terms of Youden's index, which could provide the most accurate judgment of detection.

TABLE IV. Command Injection Comparision based on Younden index.

| Tool | TP | FN | TN | FP | Total | TPR | FPR | Score |
|---|---|---|---|---|---|---|---|---|
| ZAP v2.12.0 | 79 | 47 | 125 | 0 | 251 | 62.70% | 0.00% | 62.70% |
| ZAP v2.13.0 | 71 | 55 | 110 | 15 | 251 | 56.35% | 12.00% | 44.35% |

- **Discussion of Findings:** Both versions of OWASP ZAP exhibit commendable performance, with a significant True Positive Rate (TPR) above 50%. Version 2.12.0 demonstrates a higher score, indicating its effectiveness in detecting this vulnerability.

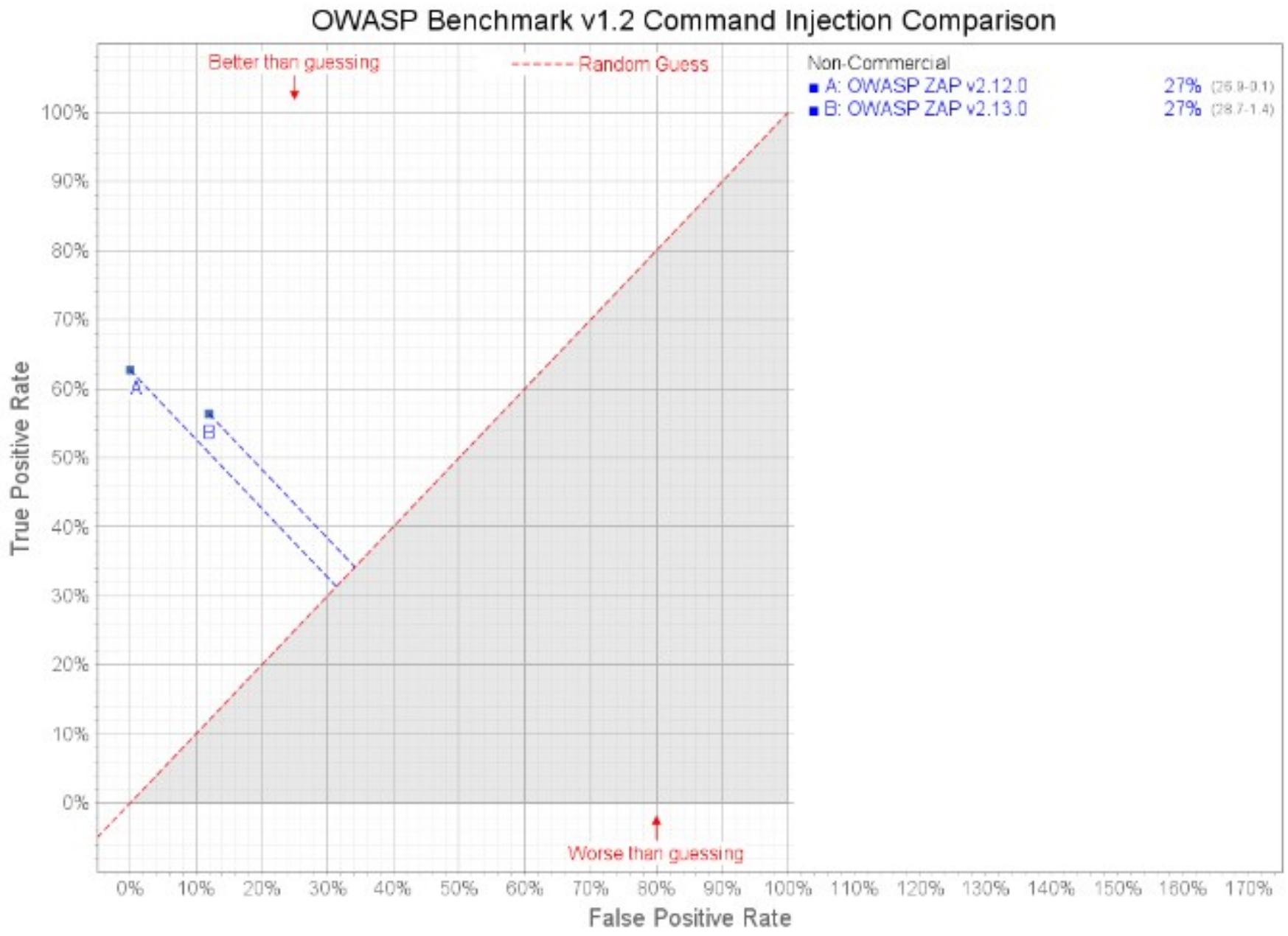

Figure 11. OWASP Benchmark v1.2 - Command Injection Comparision

**2) Path Traversal**

- **Comparison of Results:** Table V is used to compare the capability of detecting the Path Traversal vulnerability between the versions of ZAP v2.12.0 and ZAP v2.13.0

TABLE V. Path Traversal Comparision based on Younden index.

| Tool | TP | FN | TN | FP | Total | TPR | FPR | Score |
|---|---|---|---|---|---|---|---|---|
| ZAP v2.12.0 | 18 | 115 | 135 | 0 | 268 | 13.53% | 0.00% | 13.53% |
| ZAP v2.13.0 | 20 | 113 | 135 | 0 | 268 | 15.04% | 0.00% | 15.04% |

- **Discussion of Findings:** For path traversal vulnerabilities, both versions present modest results, with a TPR around 15%. This suggests that there is room for improvement in detecting these issues.

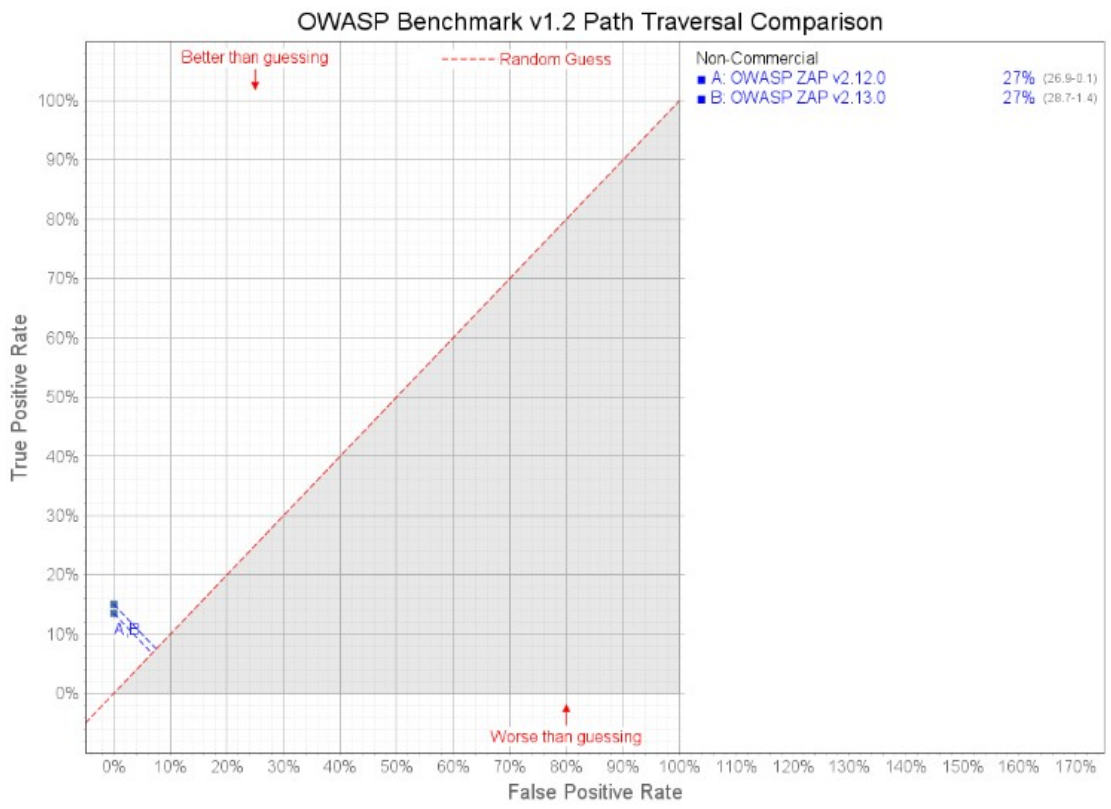

Figure 12. OWASP Benchmark v1.2 - Path Traversal Comparision

**3) Secure Cookie Flag**

- **Comparison of Results:** Table VI is used to compare the capability of detecting the Secure Cookie Flag vulnerability between the versions of ZAP v2.12.0 and ZAP v2.13.0

TABLE VI. Secure Cookie Flag Comparision based on Younden index.

| Tool | TP | FN | TN | FP | Total | TPR | FPR | Score |
|---|---|---|---|---|---|---|---|---|
| ZAP v2.12.0 | 23 | 30 | 31 | 0 | 67 | 63.89% | 0.00% | 63.89% |
| ZAP v2.13.0 | 34 | 2 | 31 | 0 | 67 | 94.44% | 0.00% | 94.44% |

- **Discussion of Findings:** The newer version, 2.13.0, significantly outperforms the previous version, achieving a higher TPR and score, indicating its enhanced ability to detect this vulnerability.

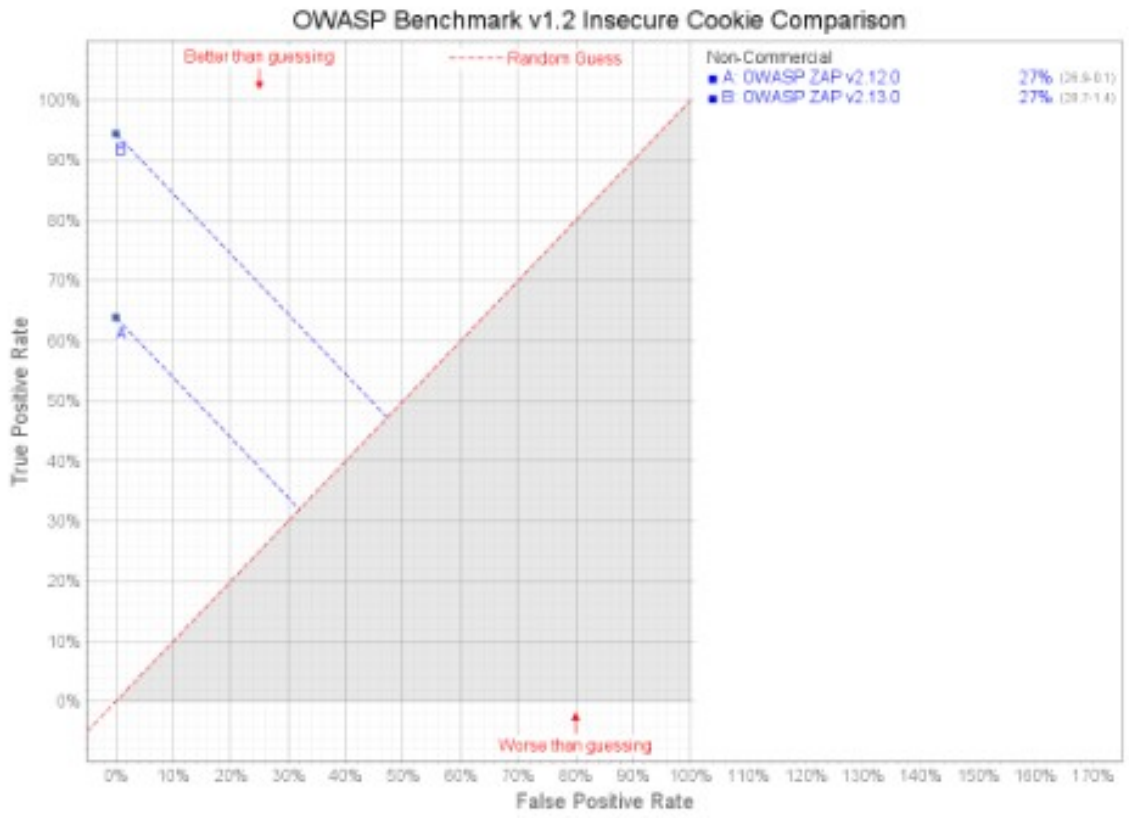

Figure 13. OWASP Benchmark v1.2 - Secure Cookie Flag Comparision

**4) SQL Injection**

- **Comparison of Results:** Table VII is used to compare the capability of detecting the SQL injection vulnerability between the versions of ZAP v2.12.0 and ZAP v2.13.0

TABLE VII. SQL Injection Comparision based on Younden index.

| Tool | TP | FN | TN | FP | Total | TPR | FPR | Score |
|---|---|---|---|---|---|---|---|---|
| ZAP v2.12.0 | 187 | 85 | 230 | 2 | 504 | 68.75% | 0.86% | 67.89% |
| ZAP v2.13.0 | 203 | 69 | 224 | 8 | 504 | 74.63% | 3.45% | 71.18% |

- **Discussion of Findings:** Version 2.13.0 surpasses its predecessor, demonstrating improved detection capabilities for SQL injection. The TPR and score are notably higher in the newer version.

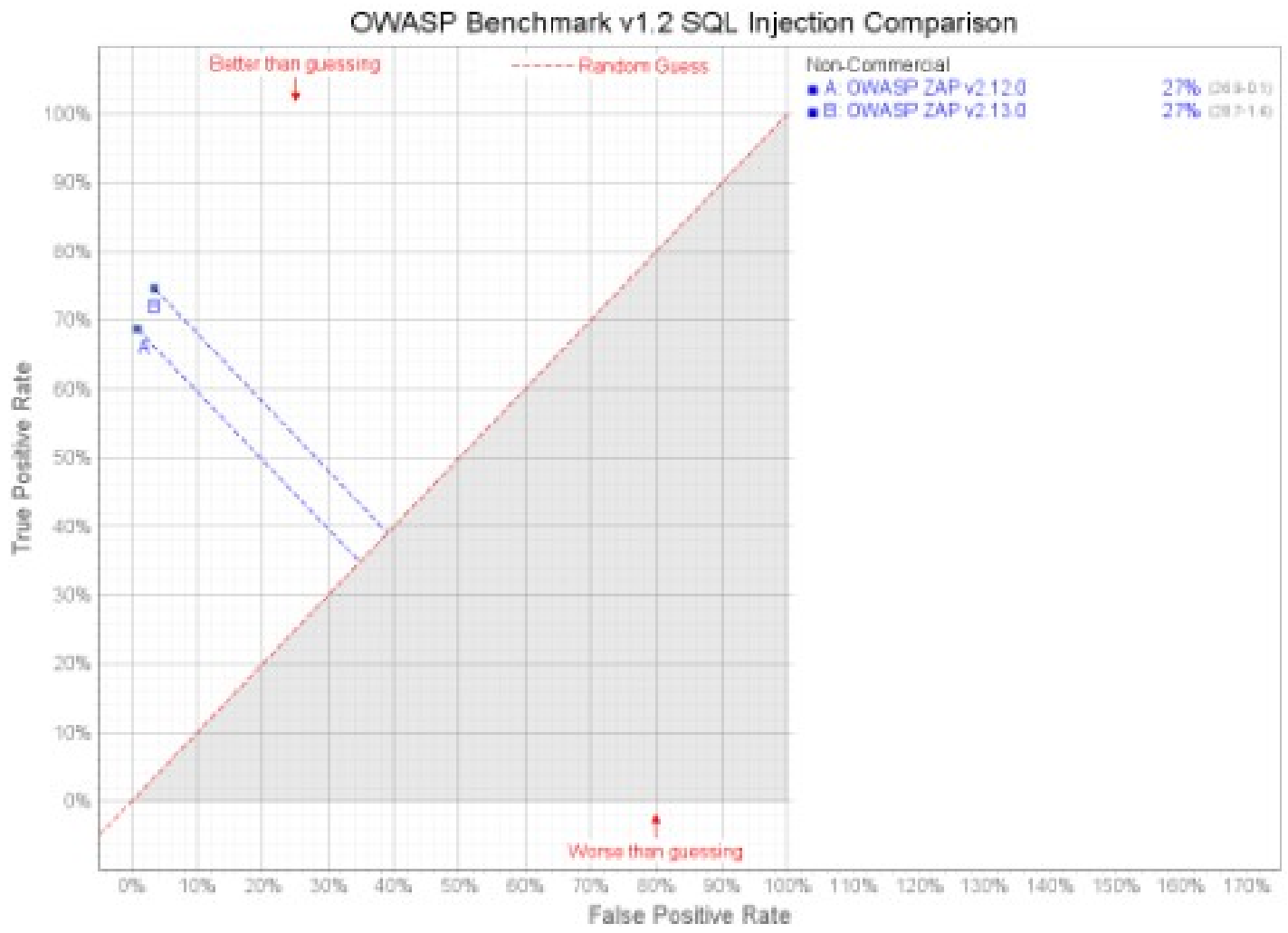

Figure 14. OWASP Benchmark v1.2 - SQL Injection Comparision

5) **XSS (Cross-Site Scripting)**

- **Comparison of Results:** Table VIII is used to compare the capability of detecting the XSS (Cross site scripting) vulnerability between the versions of ZAP v2.12.0 and ZAP v2.13.0.

TABLE VIII. XSS (Cross site scripting) Comparision based on Younden index.

| Tool | TP | FN | TN | FP | Total | TPR | FPR | Score |
|---|---|---|---|---|---|---|---|---|
| ZAP v2.12.0 | 214 | 32 | 209 | 0 | 455 | 86.99% | 0.00% | 86.99% |
| ZAP v2.13.0 | 186 | 60 | 209 | 0 | 455 | 75.61% | 0.00% | 75.61% |

- **Discussion of Findings:** In the case of XSS vulnerabilities, version 2.12.0 exhibits a higher TPR, outperforming version 2.13.0. However, the latter still maintains a respectable TPR.

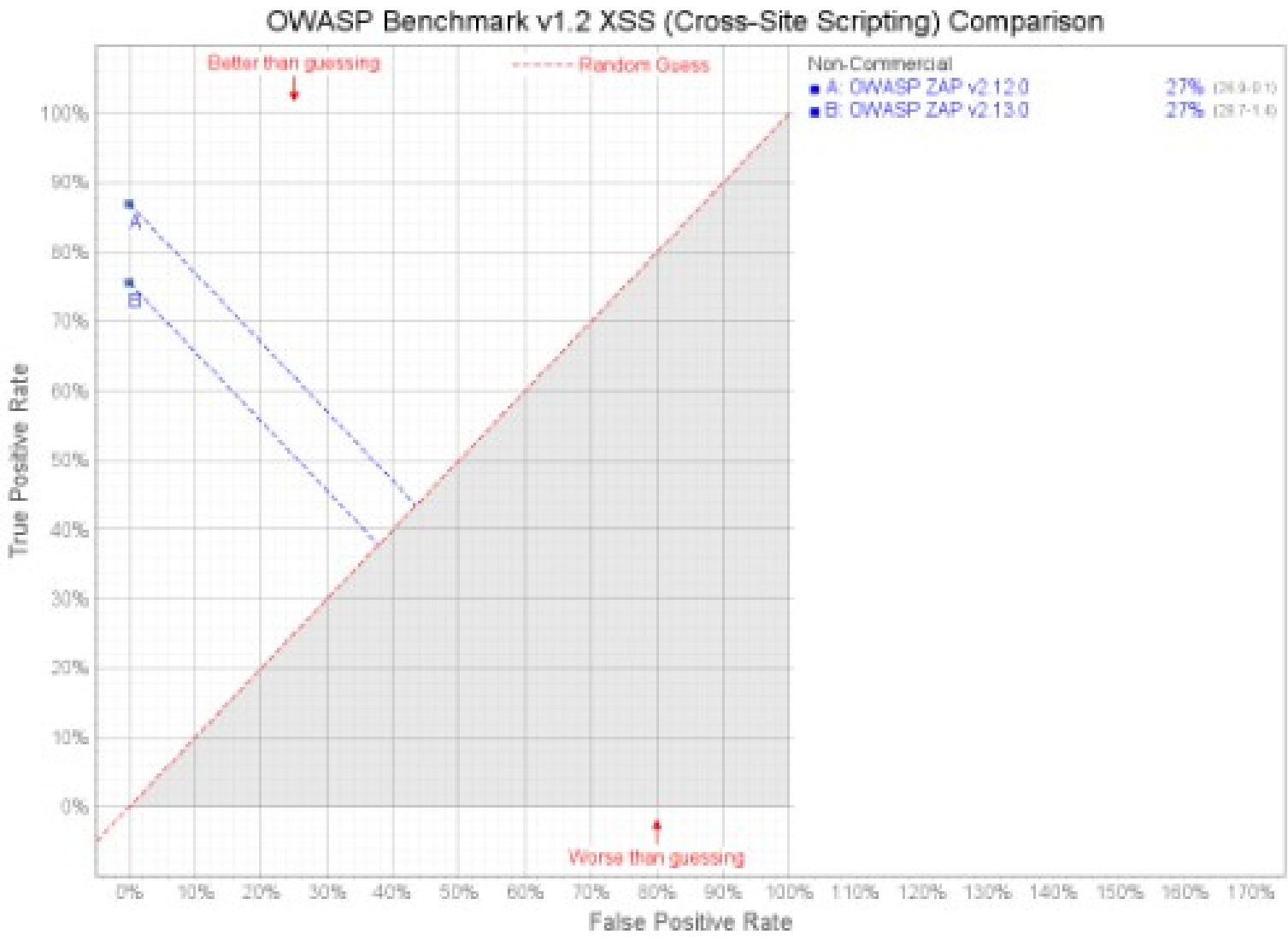

Figure 15. OWASP Benchmark v1.2 - XSS (Cross site Scripting) Comparision

C. Overall Results for V2.12.0:

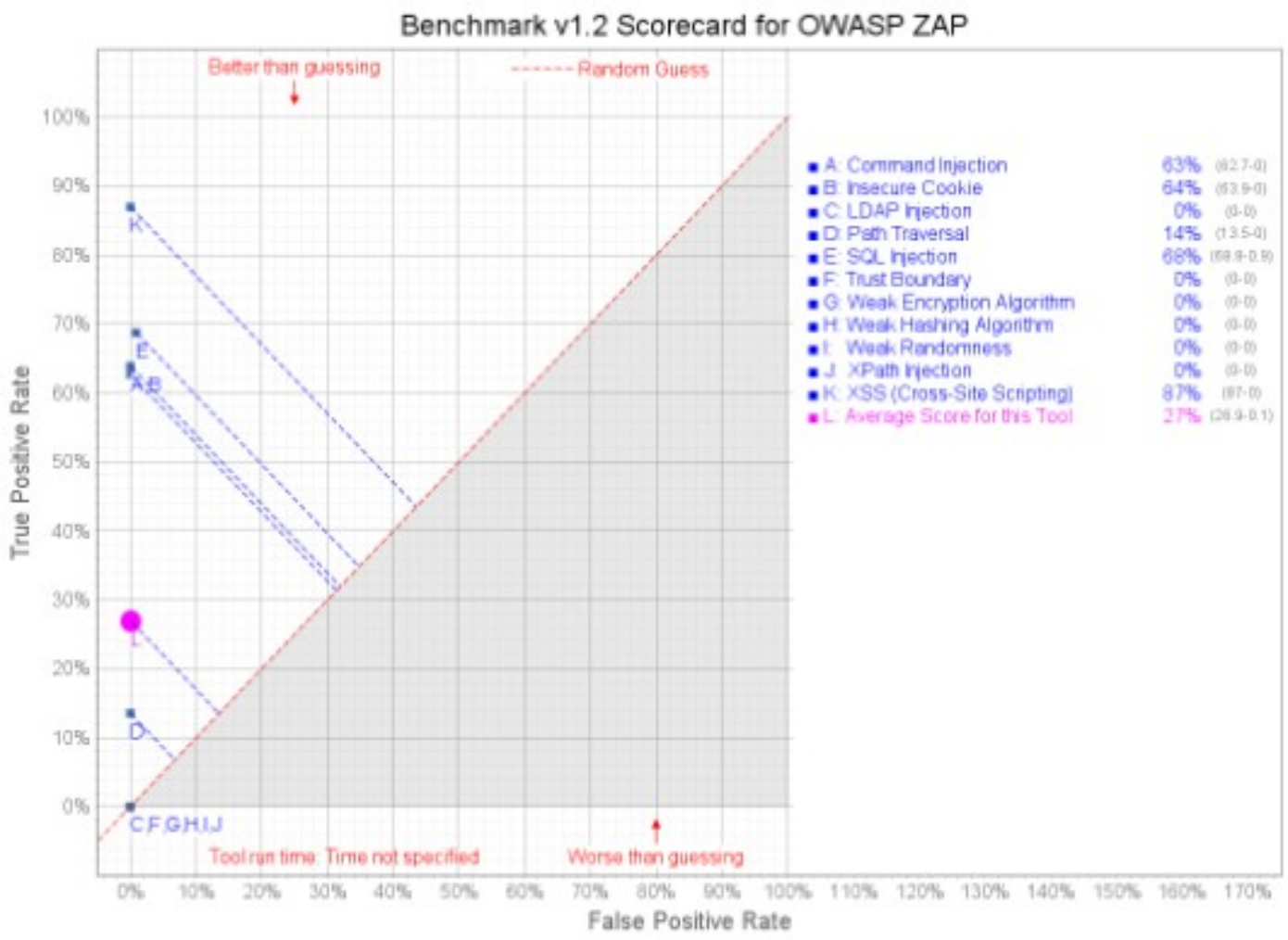

Figure 16. Benchmark v1.2 Scorecard for ZAP v2.12.0

D. Overall Results for V2.13.0:

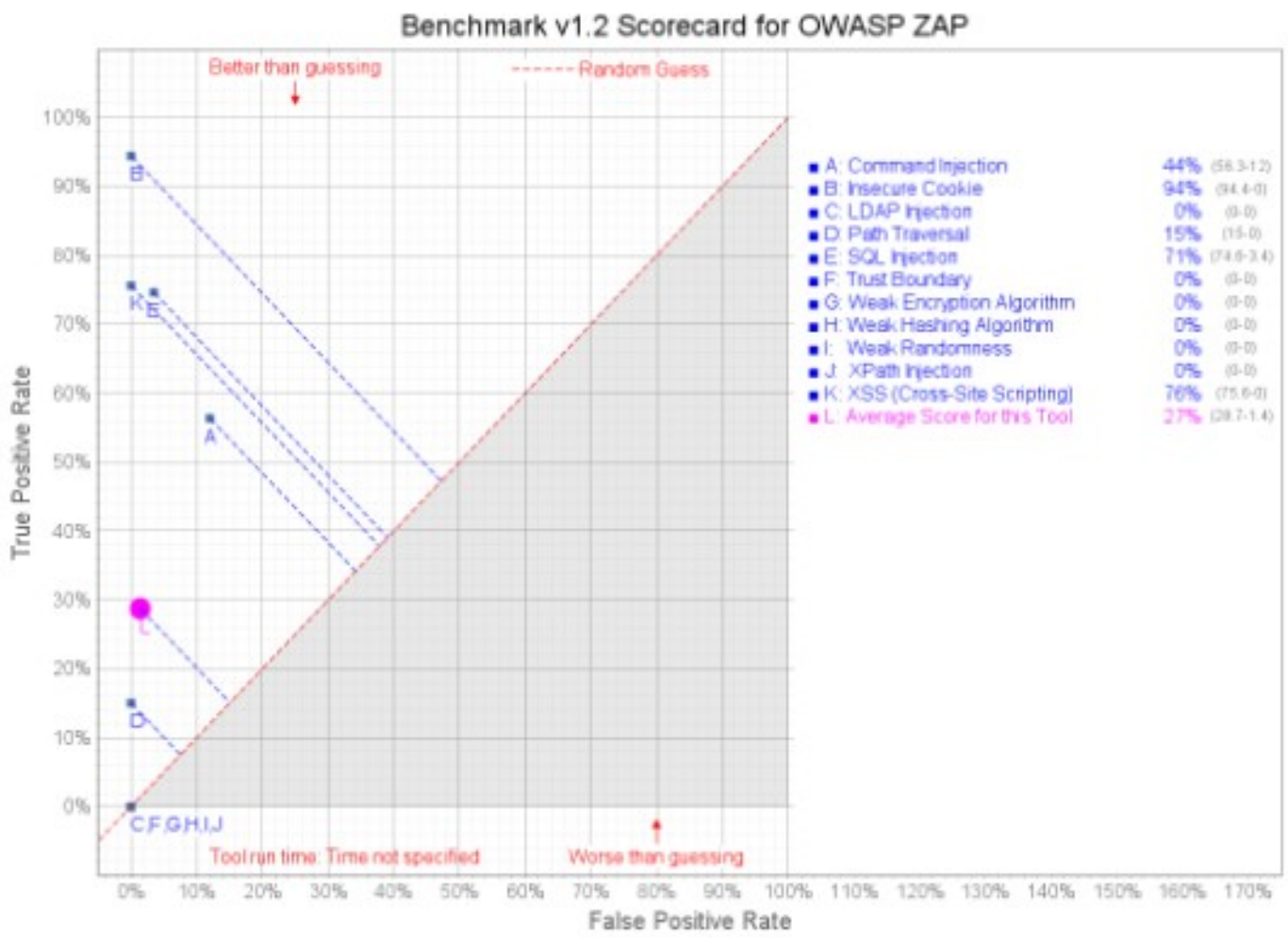

Figure 17. Benchmark v1.2 Scorecard for ZAP v2.13.0

E. Overall comparison of ZAP v2.12.0 and v2.13.0:

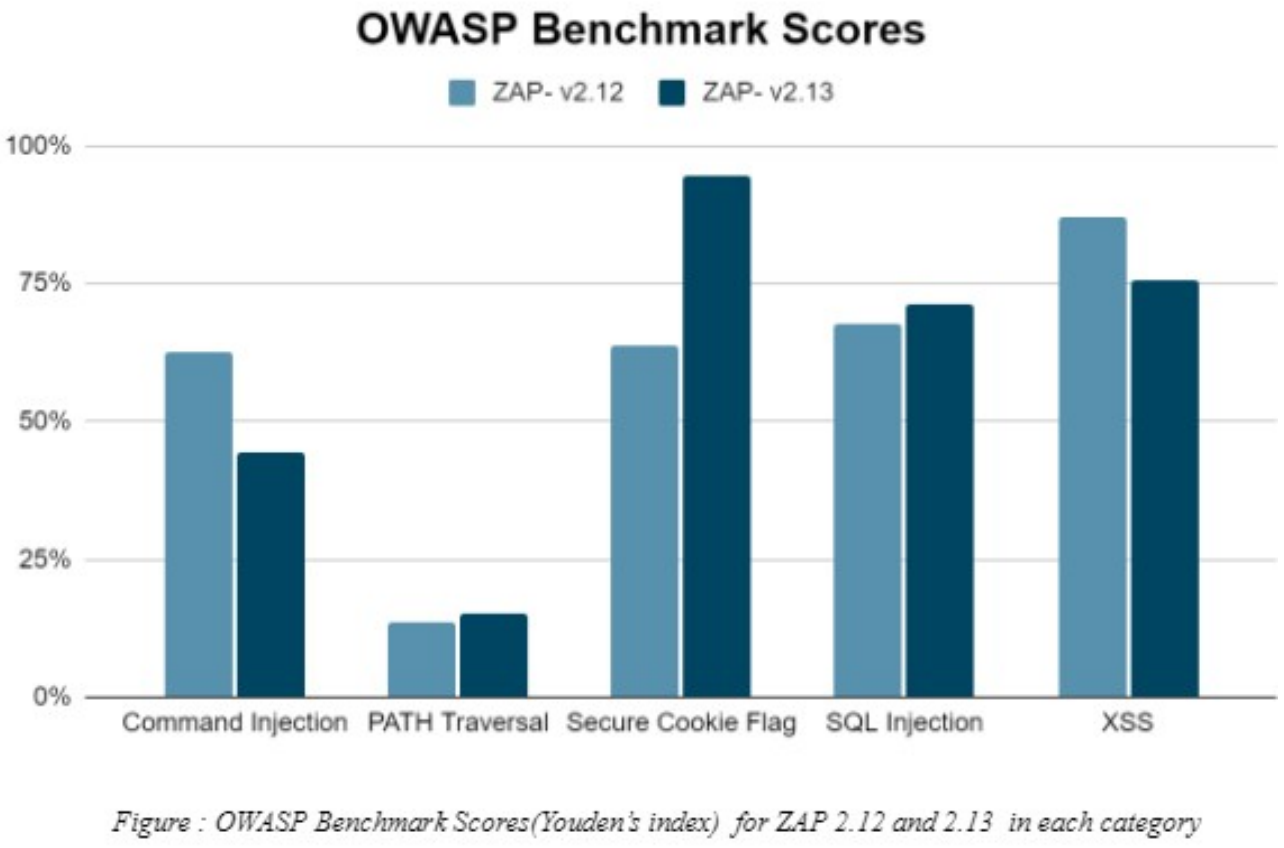

Figure 18. OWASP Benchmark Scores (Youden's index) for ZAP 2.12 and 2.13 in each category

F. Summary:

The study reveals that the choice of OWASP ZAP, a dynamic application security testing tool, plays a crucial role in determining the detection of vulnerabilities across diverse categories. While both versions, 2.12.0 and 2.13.0, are strong at identifying certain flaws, it is evident that the more recent version, 2.13.0, offers improvements in identifying secure cookie flags and SQL injection flaws. This emphasizes how crucial it is to maintain security technologies.

Ultimately, the choice of the suitable OWASP ZAP version depends on the particular security requirements of the concerned web application. Organizations can choose the best tools and versions to strengthen their web application security posture by doing this study.

The outcomes highlight the necessity of ongoing testing and development to successfully defend online applications against changing cyber threats. This study also emphasizes the importance of using the OWASP Benchmark as a powerful evaluation methodology to evaluate the effectiveness of security tools.

This study offers insightful contributions to web application security and acts as a resource for businesses looking to strengthen their security protocols.

## Conclusion

A. Conclusion

In this study, A thorough evaluation of the performance of two OWASP ZAP versions —v2.12.0 and v2.13.0 in identifying various web application vulnerabilities is carried out. The major goals of this study were to assess these versions' performance using the OWASP Benchmark project and to offer information about how well they could identify typical web application vulnerabilities. A systematic methodology that involved using OWASP ZAP as a Dynamic Application Security Testing (DAST) tool to scan web applications for vulnerabilities is employed.

- **Findings:**
  - This research covered five key vulnerability categories: Command Injection, Path Traversal, Secure Cookie Flag, SQL Injection, and XSS (Cross-Site Scripting).
  - The results revealed that both OWASP ZAP versions, v2.12.0 and v2.13.0, demonstrated varying degrees of effectiveness in detecting vulnerabilities across these categories.
  - When examining the performance of these versions, we considered key metrics such as True Positive Rate (TPR), False Positive Rate (FPR), and the calculated Youden's Index (YI) as a combined evaluation measure.
  - Across all categories, observed differences in the ability of each version to accurately identify vulnerabilities, with v2.13.0 consistently showing better performance.
- **Implications:**
  - The study highlighted the importance of using effective web application security tools, such as OWASP ZAP, in assessing vulnerabilities in a diverse range of categories.
  - The comparative analysis between the two versions demonstrated the ongoing development and improvement of OWASP ZAP.

- For security professionals, developers, and companies trying to improve their web application security testing processes, these findings offer insightful information.

This study lays the groundwork for future investigation in the area of web application security and vulnerability analysis. It is important to keep in mind that the efficiency of web application security technologies can vary based on the unique context and target apps, even though the results shown here offer insightful information. It is advised that business experts and development teams weigh the consequences of this research and change their security assessment methods as necessary.

This summary encapsulates the key outcomes of your research, highlighting the main objectives, methodologies, and findings. It provides a concise overview of your study's significance and impact on web application security assessment.

B. Future work

While this study has provided valuable insights into web application security and the comparative analysis of OWASP ZAP versions, several avenues for future research in this field are worth exploring:

- **Hybrid Testing Strategies:** Investigate the benefits of combining both Static Application Security Testing (SAST) and Dynamic Application Security Testing (DAST) techniques to provide a more comprehensive security assessment. Hybrid testing strategies could capitalize on the strengths of each approach to achieve a more robust vulnerability identification process.
- **Intelligent Tool Selection:** Develop automated methods for selecting the most suitable security tool based on the specific vulnerabilities found within a web application. Such a system could enhance the efficiency of web application security assessments by matching the tool's capabilities to the web application's security requirements.
- **Machine Learning in Vulnerability Detection:** Explore the integration of machine learning algorithms to identify and predict web application vulnerabilities. This research area could focus on developing predictive models to prevent vulnerabilities and aid in real-time threat mitigation.
- **Mobile Application Security:** Extend the scope of vulnerability assessment to mobile applications, given the increasing reliance on mobile technology. Research could delve into the adaptation of security tools like OWASP ZAP to assess the unique vulnerabilities of mobile apps.
- **Continuous Security Testing:** Investigate the implications and feasibility of incorporating continuous security testing as part of the software development lifecycle (SDLC). Determine how this approach can maintain application security and minimize the window of exposure to vulnerabilities.
- **Benchmarking Evolved Attacks:** Develop benchmarks that reflect emerging and sophisticated attack vectors. These benchmarks should challenge security tools with more advanced and real-world attack scenarios to ensure their efficacy in modern threat landscapes.
- **User-Centric Vulnerability Assessment:** Shift the focus from purely technical vulnerabilities to user-centric vulnerabilities, such as those affecting user privacy and data protection. Research could aim to assess the impact of vulnerabilities on end-users and their data.
- **Automated Vulnerability Remediation:** Investigate methods to automate the remediation of identified vulnerabilities in web applications. This research could focus on developing tools that not only detect but also suggest and implement fixes for security issues.

- **Standardization of Reporting Metrics:** Explore the standardization of reporting metrics for security tools, enabling more consistent and meaningful comparisons between various tools. Developing a universal set of assessment metrics could further improve the benchmarking process.

These future research directions aim to address emerging challenges in the dynamic field of web application security, with a particular focus on enhancing security assessment methodologies, promoting proactive vulnerability management, and optimizing the selection of security tools. By delving into these areas, researchers can contribute to the ongoing evolution of web application security practices and toolsets.

## Acknowledgments

This work was supported in part by the Ministry of Science and Technology, Taiwan, under Project MOST-110-2221-E-007-040-MY3 and MOST-111-2221- E-007-078-MY3.

## References

[1] A. Van Rensburg. Vulnerability testing in the web application development cycle. University of Johannesburg (South Africa), 2017.

[2] L. Dencheva. “Comparative analysis of Static application security testing (SAST) and Dynamic application security testing (DAST) by using open-source web application penetration testing tools,” Master thesis, dissertation, Dublin, National College of Ireland, 2022.

[3] H. S. Abdullah, “Evaluation of open source web application vulnerability scanners,” Academic Journal of Nawroz University, vol. 9, no. 1, pp. 47–52, 2020.

[4] A. Ahamed, N. Sadman, T. A. Khan, M. I. Hannan, F. Sadia, and M. Hasan, “Automated testing: Testing top 10 owasp vulnerabilities of government web applications in bangladesh,” ICSEA 2022, p. 56, 2022.

[5] A. Al Anhar and Y. Suryanto, “Evaluation of web application vulnerability scanner for modern web application,” in 2021 International Conference on Artificial Intelligence and Computer Science Technology (ICAICST). IEEE, 2021, pp. 200–204.

[6] M. Albahar, D. Alansari, and A. Jurcut, “An empirical comparison of pen-testing tools for detecting web app vulnerabilities,” Electronics, vol. 11, no. 19, p. 2991, 2022.

[7] R. Amankwah, J. Chen, P. K. Kudjo, and D. Towey, “An empirical comparison of commercial and open-source web vulnerability scanners,” Software: Practice and Experience, vol. 50, no. 9, pp. 1842–1857, 2020.

[8] K. Dasun, “A study on effectiveness of software vulnerability assessment for component-based software development,” Ph.D. dissertation, 2016.

[9] V. Di Stasio, “Evaluation of static security analysis tools on open source distributed applications,” Ph.D. dissertation, Politecnico di Torino, 2022.

[10] M. El, E. McMahon, S. Samtani, M. Patton, and H. Chen, “Benchmarking vulnerability scanners: An experiment on scada devices and scientific instruments,” in 2017 IEEE International Conference on Intelligence and Security Informatics (ISI). IEEE, 2017, pp. 83–88.

[11] J. R. B. Higuera, J. B. Higuera, J. A. S. Montalvo, J. C. Villalba, and J. J. N. P´erez, “Benchmarking approach to compare web applications static analysis tools detecting

owasp top ten security vulnerabilities." Computers, Materials & Continua, vol. 64, no. 3, 2020.
[12] C. Kalaani, "Owasp zap vs snort for sqli vulnerability scanning," 2023.
[13] E. Lavens, P. Philippaerts, and W. Joosen, "A quantitative assessment of the detection performance of web vulnerability scanners," in Proceedings of the 17th International Conference on Availability, Reliability and Security, 2022, pp. 1–10.
[14] Y. Makino and V. Klyuev, "Evaluation of web vulnerability scanners," in 2015 IEEE 8th International Conference on Intelligent Data Acquisition and Advanced Computing Systems: Technology and Applications (IDAACS), vol. 1. IEEE, 2015, pp. 399–402.
[15] B. Mburano and W. Si, "Evaluation of web vulnerability scanners based on owasp benchmark," in 2018 26th International Conference on Systems Engineering (ICSEng). IEEE, 2018, pp. 1–6.
[16] A. Chouhan, A. Halgekar, A. Rao, D. Khankhoje, and M. Narvekar, "Sentiment analysis of twitch. tv livestream messages using machine learning methods," in 2021 fourth international conference on electrical, computer and communication technologies (ICECCT). IEEE, 2021, pp. 1–5.
[17] M. Sokolova, N. Japkowicz, and S. Szpakowicz, "Beyond accuracy, f-score and roc: a family of discriminant measures for performance evaluation," in Australasian joint conference on artificial intelligence. Springer, 2006, pp. 1015–1021.
[18] A. Thakkar and R. Lohiya, "A survey on intrusion detection system: feature selection, model, performance measures, application perspective, challenges, and future research directions," Artificial Intelligence Review, vol. 55, no. 1, pp. 453–563, 2022.
[19] Ž. Vujović et al., "Classification model evaluation metrics," International Journal of Advanced Computer Science and Applications, vol. 12, no. 6, pp. 599–606, 2021.
[20] M. D. Ruopp, N. J. Perkins, B. W. Whitcomb, and E. F. Schisterman, "Youden index and optimal cut-point estimated from observations affected by a lower limit of detection," Biometrical Journal: Journal of Mathematical Methods in Biosciences, vol. 50, no. 3, pp. 419–430, 2008.
[21] "OWASP Benchmark owasp benchmark project," https://owasp.org/www-project-benchmark/, accessed: 2023-09-27.